\documentclass{revtex4}

\usepackage{graphicx}
\usepackage{amsmath}

\def\giorno{14/7/2015}

\def\a{\alpha}
\def\b{\beta}

\def\ga{\gamma}
\def\de{\delta}   
\def\eps{\varepsilon}
\def\phi{\varphi}
\def\la{\lambda}

\def\s{\sigma}
\def\z{\zeta}

\def\vth{\vartheta}

\def\L{\mathcal{L}}

\def\P{\mathcal{P}}
\def\Q{{\cal Q}}
\def\R{{\bf R}}

\def\Ga{\Gamma}

\def\La{\Lambda}
\def\Om{\Omega}

\def\pa{\partial}

\def\xb{{\bf x}}

\def\o+{\oplus}

\def\grad{\nabla}     
\def\ss{\subset}
\def\sse{\subseteq}

\def\<{\langle}
\def\>{\rangle}

\def\({\left(}
\def\){\right)}
\def\[{\left[}
\def\]{\right]}
\def\=#1{\bar #1}
\def\~#1{\widetilde #1}
\def\wt#1{\widetilde #1}
\def\.#1{\dot #1}
\def\^#1{\widehat #1}
\def\"#1{\ddot #1}

\def\eeq{\end{equation}}
\def\beql#1{\begin{equation} \label{#1}}

\def\eqref#1{(\ref{#1})}

\begin{document}

\title{Poincar\'e-like approach to Landau theory. \\ I: General theory}

\author{Giuseppe Gaeta}
\email{giuseppe.gaeta@unimi.it}
\affiliation{Dipartimento di Matematica, Universit\`a degli Studi di Milano, via Saldini 50,
I-20133 Milano (Italy)}

\date{\giorno}

\begin{abstract}
\noindent We discuss a procedure to simplify the Landau potential,
based on Michel's reduction to orbit space and Poincar\'e
normalization procedure; and illustrate it by concrete examples.
The method makes use, as in Poincar\'e theory, of a chain of
near-identity coordinate transformations with homogeneous
generating functions; using Michel's insight, one can work in
orbit space. It is shown that it is possible to control the choice
of generating functions so to obtain a (in many cases,
substantial) simplification of the Landau polynomial, including a
reduction of the parameters it depends on. Several examples are
considered in detail.
\end{abstract}



\maketitle

\section*{Introduction.}

Landau theory \cite{L1,Lan5} is a standard tool in analyzing phase
transition; it describes the state of a physical system in terms
of the minima of a certain (Landau) potential. The Landau
potential can however be very complicate, and it is thus essential
to be able to simplify it in concrete applications.

Our motivation here is mainly in applications to liquid crystals;
these will be treated more specifically in a companion paper
\cite{GLC13}, while the present one deals with the general theory.

We will assume the order $N$ at which the expansion can be
truncated is determined (we will then disregard all terms of
degree higher than $N$); this is usually done by requiring
thermodynamic stability of the resulting model. We focus then on
the problem of identifying the terms of order smaller than or
equal to $N$ which are ``inessential'', i.e. that can be dropped
without changing the qualitative properties of the Landau
potential $\Phi$. More precisely, we discuss which terms can be
eliminated by a careful choice of the coordinates in the order
parameter space; our procedure is thus fully algorithmic, requires
only to solve linear equations, and can easily be
computer-implemented via an algebraic manipulation language.

It should be stressed that our procedure, while simplifying terms
of order lower than $N$, will at the same time generate terms of
higher orders. Coherently with the general framework of Landau
theory, which considers truncated series expansions, we will not
consider these. In other words, all of our series will be
truncated at the same order $N$.

Our approach will be through application of Poincar\'e
normalization technique (see e.g. \cite{ArnG,Elp,CGs,Gae02}); we
generalize previous work \cite{AOP} in which we gave full
justification to a criterion stated by Gufan \cite{Guf,SGU} and
extended it to consider a full range of order parameters, in
particular near a phase transition.

In this paper, a more detailed analysis (compared to \cite{AOP})
of the relevant operators acting in the reduction process allows
for a more complete characterization of the reduction procedure
and of the reduced potential obtained in this way. In a companion
paper \cite{GLC13}, as already mentioned, we apply our method to
nematic liquid crystals.

It should be stressed that in the present paper we will work at
first nontrivial order, in particular when describing the effects
of the change of variables to be considered; this corresponds to
classical Poincar\'e-Birkhoff theory \cite{ArnG,Elp,CGs,Gae02}. In
the framework of Dynamical Systems, extensions of this classical
theory have been considered in order to take into account higher
order effects; this goes under the name of ``further
normalization'' \cite{Gae02} and would require some more delicate
discussion. This approach -- and its possible extension to Landau
theory -- will not be discussed here; for the application of the
Poincar\'e approach to Landau-deGennes theory of nematic liquid
crystals considered in the companion paper \cite{GLC13} it would
be rather convenient to take into account these higher order
effects as well. \footnote{In that paper they will however be
considered via a ``brute force'' approach; this will avoid to
enter into the mathematical details needed for further
normalization, which are therefore not discussed here.}

The main result of the paper will be the formulation of an
algorithmic method (requiring only the solution of linear
equations) to simplify the Landau potential. It should be stressed
that this will be just based on mathematical manipulations, and
more specifically on the choice of adapted coordinates (which are
built through a perturbation approach); that is, we will not
introduce any physical considerations. Introducing the latter
could of course result, in principles, in a further reduction of
the Landau potential. Thus our procedure should be seen as a first
step eliminating unnecessary complications through the choice of
convenient coordinates; it leaves room for further reduction based
on the Physics of the specific system under study.

We also anticipate that in many cases the resulting simplified
potential will still be quite complex, despite a substantial
reduction. E.g., in examples 5 and 6 below, we will have to
consider a potential depending in principles on 22 parameters, and
will be able to eliminate 16 of them; but the simplified potential
will still depend on 6 parameters and thus be very hard to study.
\bigskip

The {\it plan of the paper} is as follows. In Sect.\ref{sec:prel}
we will recall some basic notions, to be used in the following,
also in order to set down our general notation. In
Sect.\ref{sec:transf} we will start applying the Poincar\'e
approach: we will identify a class of changes of coordinates which
is convenient for our purposes; these are identified by generating
functions which are the gradients of invariant functions. We also
discuss how invariant polynomials are transformed under these. In
Sect.\ref{sec:red} we will apply this discussion to Landau
polynomials, and see how one can choose the generating functions
mentioned above in order to obtain a simpler Landau polynomial.
Section \ref{sec:Examples} is then devoted to illustrate the
different steps analyzed theoretically in previous sections by
means of a number of case studies. In Sect.\ref{sec:adapt} we
discuss how the analysis can be simplified by passing to adapted
coordinates; and in Sect.\ref{sec:reduction} how the reduction
procedure can be of help in analyzing the behavior of physical
systems, i.e. in performing a -- quantitative and qualitative --
analysis of Landau polynomials. Finally, in Sect.\ref{sec:disc} we
discuss the advantages and the limitations of our methods, as well
as possible direction of further extension of the present results;
and in the brief Sect.\ref{sec:concl} we draw some conclusions.

\section{Preliminaries}
\label{sec:prel}

We will briefly recall some notions from Landau theory, mainly to
set the notation to be used below; see also \cite{AOP} for details
and for further references (in particular on invariants theory).
Summation over repeated indices is understood.

\subsection{Generalities}

We denote by $x \in M \sse \R^m$ the order parameter, and by $G$
the group acting in $M$ to describe symmetry of the system in the
order parameters space; this acts through a real representation,
i.e. a set of matrices $\{ T_g , g \in G \}$; as this is fixed and
in order to avoid cumbersome notation, we also just write $g$ for
$T_g$ as well. We assume $G$ is compact, and it acts linearly and
orthogonally in $\R^m$, mapping the order parameter space $M$ to
itself. \footnote{ Note that if $G$ is a continuous group, some of
the $x$ variables will actually be inessential physically, as can
be quotiented out; in the present note we only consider discrete
groups, while in the companion paper \cite{GLC13} we will meet
this situation.}

The effective potential $\^\Phi (x) \in \R$ is a $G$-invariant
polynomial, so we should preliminarily determine the most general
$G$-invariant polynomial in the $x^i$; the Landau potential $\Phi
(x)$ will be a truncation of $\^\Phi$ to a suitable order $N$. We
would then like to omit some ``unessential'' terms of order $n \le
N$ in $\Phi$ (see below).

The polynomials $\^\Phi$ and $\Phi$ have coefficients depending on
external (control) parameters $\la \in \La \sse \R^L$, so that we
will also write $\^\Phi (x;\la)$, $\Phi(x;\la)$; the equilibrium state of the
system is described by the minima of $\Phi (x;\la)$, which we
denote as $x_\a (\la)$. In general there will be different minima
for a given value of $\la$: in particular, if $x (\la)$ is not a
fixed point for the $G$-action, then the whole $G$ orbit through
$x (\la)$ will be an orbit of minima. Moreover, there can be
different $G$-orbits of minima for a given value of $\la$.

The symmetry of the state corresponding to $x (\la)$ will
correspond to $G_{x(\la)}$, the isotropy group of $x (\la)$; we
recall that by definition $G_x := \{ g \in G :  T_g x = x\}$.

If $x$ and $y$ are on the same $G$-orbit, $y = g x$, then $G_y = g G_x
g^{-1}$; the conjugacy class of isotropy subgroups associated to
any $G$-orbit in $M$ is called the {\it orbit type} $[Gx]$. A {\it
phase} will be described by an orbit type.

A necessary condition to have a phase transition at $\la = \la_0$
is that the orbit type $[G x(\la)]$ is not constant in a
neighborhood of $\la_0$, no matter how small.

We stress again that it is inherent to Landau theory to consider a
truncated series expansion; thus all of our computations and
theoretical considerations will disregard higher order terms. In
particular, when our change of variables would generate terms of
order higher than $N$, these will simply be dropped.

\subsection{Invariant polynomials}

By the {\it Hilbert basis theorem} \cite{AbS,PrS}, there is a set
$\{J_1 (x) , ... , J_r (x) \}$ (with our hypotheses on $G$, $r$ is
guaranteed to be finite \cite{AbS}) of $G$-invariant homogeneous
polynomials of degrees $\{d_1 , ... , d_r \}$ (we can and will
always order these so that $d_1 \le d_2 \le ... \le d_r $) such
that any $G$-invariant polynomial $\^\Phi (x)$ can be written as a
polynomial in the $\{ J_1 , ... , J_r \}$, i.e. \beql{1.1} \^\Phi
(x) \ = \ \^\Psi \[ J_1 (x) , ... , J_r (x) \] \eeq with $\^\Psi$
a polynomial in $(J_1,...,J_r)$.

When the $J_a$ are chosen so that none of them can be written as a
polynomial of the others and $r$ has the smallest possible value,
we say that they are a {\it minimal integrity basis (MIB)}, and
that the $\{J_a \}$ are a set of {\it basic invariants} for $G$.

When the elements of a MIB for $G$ are algebraically independent,
we say that the MIB is {\it regular}; not all groups $G$ admit a
regular MIB (see example 2 below).

We will from now on assume we have chosen a MIB, with elements
$\{J_1 , ... , J_r\}$ (of degrees $\{d_1 , ... , d_r \}$ in $x$,
with $d_1 \le d_2 \le ... \le d_r$).

\subsection{The (Sartori) $\P$-matrix}

In the following we will need to consider a matrix built with the
gradients of basic invariants, which we call the Sartori {\it
$\P$-matrix} \cite{Sar}. This is defined, with $\< .,.\>$ the
standard scalar product in $M = \R^m$, as \beql{1.2} \P_{ih} (x) \
:= \ \< \grad J_i (x) , \grad J_h (x) \> \ . \eeq The gradient of
an invariant is necessarily a covariant quantity; the scalar
product of two covariant quantities is an invariant one, and thus
can be expressed again in terms of the basic invariants. Moreover,
we always deal with polynomials. Thus, {\it the $\P$-matrix can
always be written in terms of the $J$ themselves}.

\subsection{Orbit space and the Michel principle}

Let us come back to $\^\Phi$; this is $G$-invariant and thus can
be written in terms of the basic invariants.  The evaluation of
$\^\Phi : M \to \R$ is in principles substituted by evaluation of
${\bf J}: M \to \Om$ and $\^\pi : \Om  \to \R$; here we have
denoted by $\Om \sse \R^r$ the target space for ${\bf J} = (J_1 ,
... , J_r)$.

If -- as in Landau theory -- we have to consider the most general
$G$-invariant polynomial on $M$, we only have to deal with the map
$\pi : \Om \to \R$. In general $\Om$ is a semi-algebraic
submanifold (i.e. it is defined by polynomial equalities and
inequalities) of $\R^r$, possibly of dimension smaller than $r$;
if the MIB is regular then $\Om$ has dimension $r$.

The space $\Om$ is also known as the {\it orbit space} for the $G$
action on $M$; indeed, its points are in one-to-one correspondence
with the $G$-orbits in $M$, $\Om \simeq M/G$. The geometry of
orbit space is discussed e.g. in \cite{AbS,Sar,Mic,MZZ}; for
applications to Dynamics, see e.g. \cite{Cho}.

The {\bf Landau-Michel principle} states that {\it Landau theory
can be worked out in the $G$-orbit space $\Om := M / G$.}

\subsection{Thermodynamic stability and convexity}
\label{sec:stability}

Let us now briefly discuss how the request of thermodynamic
stability \cite{Lan5}, i.e. convexity, is reflected in the polynomial $\^\pi
(J)$.

Consider first the regular case; now $\^\pi : \R^r \to \R$,
and the $J_a$ can be considered as independent variables. The
minimal Landau polynomial $\Phi (x) = \pi (J)$ will be quadratic
in the $J$, and the stability is ensured by requiring that the
matrix $D_{ih} = \pa^2 \Phi / (\pa x^i \pa x^h )$ is positive
definite for $|x|$ sufficiently large.

So the prescription in this case will be to {\it consider a
polynomial of order (at least) $N = 2 \, {\rm max} (d_1,...,d_r) =
2 d_r$}; and of course choose coefficients so that the matrix $D$
is positive definite for large $|x|$.

If we deal with a non-regular case, this prescription also works:
maybe it would also be possible to stop at a lower order, as we
have to care only about the submanifold of $\Om$ allowed by the
relations between the $J_a$, but if we require stability in all of
$\Om$ we are on the safe side.

Some remarks are in order here:
\begin{itemize}
\item[(i)] The prescription is {\it not } to write $\pi$ as a quadratic
polynomial in the $J_a$ and then express $\Phi$ in terms of this;
rather it is to {\it consider the most general $G$-invariant
polynomial of order $2 d_r$} (this can contain quite high powers
in some of the $J_a$'s, see examples below).
\item[(ii)] The requirement to have $D$ positive definite for large $|x|$
is surely satisfied if the largest order term in $\Phi (x)$ is a
power of $\rho = |x|^2$.
\item[(iii)] The coefficients of (at least some of) the polynomials
will depend on the external parameters; in particular, this will
be the case for $J_1 = |x|^2$, whose coefficient controls the loss
of stability of the critical point $x = 0$ and thus the onset of
the phase transition.
\end{itemize}

\section{Transformation of invariant polynomials}
\label{sec:transf}

We will apply to Landau theory the technique of {\it Poincar\'e
transformations}. These are the fundamental tool of the theory of
Poincar\'e-Birkhoff normal forms \cite{ArnG,CGs,Elp}; see
\cite{AOP} for a more complete discussion of the relation of these
with Landau theory.

Let us consider $G$-invariant polynomials $\Phi (x) = \Psi [J (x)]$.
We write \beql{eq:2.1} \Phi (x) \ = \ \sum_{k=0}^\infty \, \Phi_k
(x) \eeq where $\Phi_k (a x) = a^{k+2} \Phi_k (x)$.

We want to consider changes of coordinates of the form
\beql{eq:2.2} x^i \ \to \ x^i \ + \ h^i_m (x) \ , \eeq with $h_m
(a x) = a^{m+1} h_m (x)$; moreover, we want to preserve the
symmetry properties of $\Phi$. Thus the function $h: M \to M$ has
to transform in the same way as  $x$ under the $G$-action, i.e. we
have to require $$ h (T_g x) \ = \ T_g \, h (x) $$ for all $x \in
M$ and all $g \in G$. We will choose $h_m$ to be the gradient of a
$G$-invariant function $H_m (x)$, i.e. $$ h^i_m (x) \ = \ g^{ij} \
(\pa H_m / \pa x^j) \ , $$ with $g$ the metric in $M$; note $H_m
(a x) = a^{m+2} H_m (x)$.

As $H_m$ is $G$-invariant, it is also possible to write it as a
function of the basic invariants: $H_m (x) = \chi_m [J_1 (x) , ...
, J_r (x)]$. This yields \beql{eq:2.2h} (\pa H_m / \pa x^i) \ = \
(\pa \chi_m / \pa J_\a) \cdot (\pa J_\a / \pa x^i) \ . \eeq

In order to know how \eqref{eq:2.2} acts on \eqref{eq:2.1}, it
suffices to know how it acts on the basic invariants $J_a$, and
how this is reflected in the action on $\Phi$. The computations
are straightforward, but the resulting formulas can be
considerably involved.

Luckily, we will only need the first order terms; dropping higher
order terms, recalling the expression for $h^i$, and using
\eqref{eq:2.2h}, we get
\begin{eqnarray}
 J_a (x) &\to& J_a (x) \ + \ (\pa J_a / \pa x^i) \de x^i \nonumber \\
 &=& J_a (x) \ + \ (\pa J_a / \pa x^i)
g^{ij} (\pa H_m / \pa x^j) \nonumber \\
 &=& J_a (x) \ + \ (\pa J_a / \pa x^i)
g^{ij} (\pa \chi_m / \pa J_b) (\pa J_b / \pa x^j) \nonumber \\
&=&  J_a (x) \ + \ \P_{ab} \, (\pa \chi_m / \pa J_b) \ .
\label{eq:2.11} \end{eqnarray}

\section{Reduction of Landau polynomials}
\label{sec:red}

Let us now apply the above discussion to the reduction of an
invariant polynomial $\Phi (x) = \Psi (J_1 , ... , J_r )$. That
is, we want to choose the $\chi_m$ so to obtain a convenient
(reduced) form of the polynomials $\Psi_k$, hence of the $\Phi_k$
as well.

\subsection{General reduction scheme}

We have in general, dropping h.o.t. as usual, \beql{eq:2.12} \Psi
(J) \ \to \ \Psi (J + \de J) \ = \ \Psi (J) + \sum_{\a=1}^r \,
\frac{\pa \Psi (J)}{\pa J_\a} \, \de J_\a \ . \eeq We have seen
that under \eqref{eq:2.2} the $J_a (x)$ change according to
\eqref{eq:2.11}, hence we readily obtain at first order
\beql{eq:2.13} \de \Psi \ = \ \frac{\pa \Psi}{\pa J_\a} \, \P_{\a
\b} \, \frac{\pa \chi_m}{\pa J_\b} \ . \eeq

Let us now consider the expansion \eqref{eq:2.1} for $\Phi$, and
write correspondingly $\Psi = \sum_k \Psi_k$, where $\Phi_k (x) =
\Psi_k [J(x)]$. We will also expand the $\P$-matrix in homogeneous
terms, $\P = \sum_k \P^{(k)}$, again with $\P^{(k)}$ homogeneous
of degree $(k+2)$ in the $x$. Note that $\P_{ih}$ is homogeneous
of degree $(d_i + d_h - 2)$; in particular, $\P^{(0)}$ corresponds
to the sub-matrix relating quadratic invariants only.

It follows from our discussion that the terms $\Psi_k$ with $k <
m$ are not changed, while the term $\Psi_m$ changes as
\beql{eq:2.14} \Psi_m \ \to \ \Psi_m \ + \ \frac{\pa \Psi_0}{\pa
J_\a} \ \P^{(0)}_{\a \b} \ \frac{\pa \chi_m}{\pa J_\b} \ . \eeq
Terms of higher order change in a more complex way.
\footnote{Actually, their transformation can be described by means
of the Baker-Campbell-Haussdorff formula \cite{CGs}; but this is
inessential here.}

We can thus operate sequentially with $H_1, H_2,H_3,...$; at each
stage (generator $H_m$) we are not affecting the terms $\Psi_k$
with $k < m$. Moreover, we can just consider the first order
correction; higher order terms will be changed in some complex way
but they were generic, hence will continue being such, and (those
of degree not higher than the truncation order $N$) will be taken
care of in subsequent steps.

It should be stressed that in \eqref{eq:2.14} it is {\it not} the
full $\P$-matrix which appears, but only its quadratic (in $x$)
part $\P^{(0)}$; this in turn only depends (linearly) on the
quadratic invariants.

\subsection{Analysis of the reduction procedure}

Let us now consider \eqref{eq:2.14} in more detail; we are
interested in the case where $x=0$ is always a critical point for
the Landau polynomial (albeit not necessarily a minimum); then
there will be invariants $J_1 , ... , J_s$ (with $s \le r$)
quadratic in the $x$, and only these will appear in $\Psi_0$,
which will be of the form \beql{eq:2.15} \Psi_0 \ = \
\sum_{\a=1}^s c^\a \ J_\a \eeq with $c^\a$ some real constants.
Thus \eqref{eq:2.14} can be rewritten as $\Psi_m \to \Psi_m + \de
\Psi_m$ with \beql{eq:2.14b} \de \Psi_m \ = \ c^\a \, \P^{(0)}_{\a
\b} \ \frac{\pa \chi_m}{\pa J_\b} \ := \ - \, \L_0 (\chi_m ) \ . \eeq It
is quite clear that any term in $\Psi_m$ which lies in the range
of the linear differential operator \beql{eq:L0} \L_0 \ := \ - \,
c^\a \, \P^{(0)}_{\a \b} \ \frac{\pa }{\pa J_\b} \eeq can be
eliminated by a suitable choice of $\chi_m$. On the other hand, we
can always add to $\chi_m$ some term in the kernel of $\L_0$
without affecting $\de \Psi_m$.

More precisely, let us denote by $S_m$ the set of smooth functions
$F : M \to \R$ which are $G$-invariant and homogeneous of degree
$(m+2)$ in the $x$; and by $\pi_m$ the operator of projection to
$\mathtt{Ran} (\L_0 ) \cap S_m$. Then any term $\Psi_m \in \pi_m
(S_m)$ can be eliminated via the step-$m$ Poincar\'e
transformation by choosing the generating function as $H_m (x) =
\chi_m [J(x)]$ with $\chi_m$ a solution to the equation
\beql{eq:hom} \L_0 (\chi_m) \ = \ \pi_m \, \Psi_m \ ; \eeq we will
refer to this as the {\it homological equation}, like in standard
Poincar\'e normal forms theory \cite{ArnG,Elp}.

It should be noted that if $\pi_m \Psi_m = \Psi_m$ the homological
equation \eqref{eq:hom} determines the generating function to
completely cancel the $\Psi_m$ term; but when we deal with the
highest order terms in the Landau polynomial ($m = N$), we do not
want to completely cancel these. In fact, we should be careful to
preserve the thermodynamical stability (i.e. the convexity at
large $|x|$, as discussed in Sect.\ref{sec:stability}), see below.

\medskip\noindent
{\bf Remark 1.} It should be stressed that the whole procedure is
based on a non-degeneration hypothesis, i.e. on the assumption to
have a non-zero quadratic part $\Psi_0$. If this is not the case,
i.e. if the $c^\a$ in \eqref{eq:2.15} are all zero, the
homological operator $\L_0$ is trivial, and the theory simply
vanishes \footnote{One could actually consider a ``higher order
normalization'' e.g. following the steps of \cite{Gae02}; but we
prefer not to enter into such details.}. Thus the reader should
not be surprised if later on, in concrete examples, he/she will
always find that results depend on conditions amounting indeed to
the non-vanishing of the quadratic part of the potential.

\subsection{The operator $\L_0$}
\def\Q{\Theta}

Let us consider in more detail the operator $\L_0$. As remarked
above, $\P^{(0)}$ necessarily depends only on the quadratic
invariants $J_1 , .... , J_s$ (with $1 \le s \le r$), and is
linear in these. Thus we can always write $\P^{(0)} = \kappa^\ga
J_\ga$ with $\kappa^\ga$ a constant real matrix (with numerical
entries $K_{\a \b}^\ga$), i.e. \beql{eq:K} \P^{(0)}_{\a \b} \ = \
K_{\a \b}^\ga \ J_\ga \ ; \eeq the real coefficients $K_{\a
\b}^\ga$ (and the matrices $\kappa^\ga$) are identically zero for
$\ga > s$.

With this notation, the operator $\L_0$ reads \beql{eq:L0K} \L_0 \
= \ - \, [ (c^\a \, K_{\a \b}^\ga ) \, J_\ga ] \  \frac{\pa }{\pa
J_\b} \ := \  - \ (Q_\b^{\ \ga} \, J_\ga ) \, \frac{\pa}{\pa J_\b}
\ . \eeq

Note that we can reach the same expression in a slightly different
way; indeed, \eqref{eq:L0} can be rewritten as \beql{eq:L0q} \L_0
\ = \  - \ \Q_{\b} \ (\pa / \pa J_\b ) \ , \eeq with of course
$\Q_{\b} = c^\a \P^{(0)}_{\a \b}$; as we know that $\P^{(0)}$ is
linear in the $J$,  necessarily $\Q_\b = Q_\b^{\ \ga} J_\ga$ for
some constant matrix $Q$, and we arrive again at \eqref{eq:L0K}.

\subsection{Normalized versus original coordinates}

It should be stressed that the simplification (or normalization,
{\it \`a la Poincar\'e}) procedure is based on passing from the
original coordinates -- which in this case are the order
parameters -- to new coordinates which are expressed as {\it
non-homogeneous} functions of the old ones.

This means in particular that albeit one may have at first sight
the impression that the reduced Landau polynomial supports phase
transitions of order different from the original one (e.g., in the
case where the next to lowest order terms are fully cancelled),
when the predictions obtained on the basis of the reduced
polynomial are mapped back to the original coordinates, one does
of course go back to the original phase transition order.

More generally, as what we do here is just a sequence of changes
of coordinates, it is clear that no physical predictions can be
altered -- albeit obtaining such predictions may be simpler in the
new coordinates.

\section{Examples}
\label{sec:Examples}

In order to fix ideas, let us consider explicitly some concrete
(simple) example. We will follow our theoretical discussion, and
consider ``in parallel'' the different steps for various examples,
in different subsections; we trust this will better help the
reader to familiarize with the present approach. We will also deal
with some of the examples considered in \cite{AOP}, in order to
make easier comparison with the methods used in previous work.

Note that the symmetry considered in Example 1 is the one of
bent-core (or chevron-shaped) nematic liquid crystals, that in
Example 4 is the one of isotropic nematics, and those of Examples
5 and 6 are relevant to anisotropic nematics \cite{GLJ,AlL}.

Here all the indices will be written as lower ones, in order to
avoid any possible confusion with exponents.

\subsection{The $\P$ and $Q$ matrices}
\label{sec:ExI}

\bigskip\noindent
{\bf Example 1.}
Consider $M = \R^2 = \{ x,y \}$ with group $G = Z_2 \times Z_2$ generated by
$$ g_x : (x,y) \to (-x,y) \ , \ \ g_y : (x,y) \to (x,-y) \ ; $$
in this case the MIB is given by two invariants, both of them quadratic:
$$ J_1 \ = \ x^2 \ , \ \ J_2 \ = \ y^2 \ . $$ Note that here $\rho := |x|^2$ is written in
terms of the chosen basic invariants as $\rho = J_1 + J_2$.

We have $\P^{(0)} = \P$, and we get immediately
$$ \P = \P^{(0)} = \begin{pmatrix}  4 J_1 & 0 \\ 0 & 4 J_2 \end{pmatrix} \ ; \ \
\Q = \underline{c} \, \P^{(0)}
 = \begin{pmatrix} 4 c_1 J_1 \\ 4 c_2 J_2 \end{pmatrix} \ . $$
In other words, now the matrix $Q$ is diagonal,
$$ Q \ = \ \begin{pmatrix} 4 c_1 & 0 \\ 0 &  4 c_2 \end{pmatrix} \ . $$

\bigskip\noindent
{\bf Example 2.} Consider $M = \R^2$, and $G = Z_2$ is the group
generated by inversion (simultaneous reflections in $x$ and in
$y$), i.e. by
$$ g : (x,y) \to (-x,-y) \ . $$
Now the MIB is given by three invariants, all of them quadratic:
$$ J_1  = x^2 \ , \ \ J_2  = y^2 \ , \ \ J_3 = x y \ . $$
Again $\rho = J_1 + J_2$.
Note that the basis is {\it not} regular: we have $J_1 J_2 = J_3^2$.

Again $\P^{(0)} = \P$, and in this case we get
{\small
$$ \P (x,y) \ = \ \begin{pmatrix} 4 x^2 & 0 & 2 x y \\ 0
& 4 y^2 & 2 x y \\ 2 x y & 2 x y & x^2 + y^2 \end{pmatrix} \ = \
\begin{pmatrix} 4 J_1 & 0 &
2 J_3 \\ 0 & 4 J_2 & 2 J_3 \\ 2 J_3 & 2 J_3 & J_1 + J_2 \end{pmatrix} \ . $$ }
It follows that
$$ \Q = \begin{pmatrix} 4 c_1 J_1 + 2 c_3 J_3 \\ 4 c_2 J_2 + 2 c_3 J_3 \\
c_3 (J_1 + J_2) + 2 c_1 J_3 + 2 c_2 J_3 \end{pmatrix} \ , $$ and hence
$$ Q \ = \ \begin{pmatrix} 4 c_1 & 0 & 2 c_3 \\ 0 & 4 c_2 & 2 c_3 \\
c_3 & c_3 & 2 (c_1 + c_2)  \end{pmatrix} \ . $$

\bigskip\noindent
{\bf Example 3.}
Consider now the group generated by rotation of an angle $\vth = (2 \pi / 3)$
in the plane $(x,y)$; with $R$ the rotation matrix
$$ R \ = \ \begin{pmatrix} \cos ( \vth ) & - \sin (\vth ) \\ \sin (\vth ) & \cos (\vth ) \end{pmatrix} \ , $$
the group consists simply of $G = \{ I , R , R^2 \}$. The basic
invariants are
$$ J_1 = x^2 + y^2 \ , \
J_2 = x^3 - 3 x y^2 \ , \ J_3 = y^3 - 3 x^2 y \ ; $$ only the
first one is quadratic \footnote{With $z = x + i y$, these
correspond to $J_1=|z|^2$, $J_2 = \mathtt{Re} [z^3]$, $J_3 = -
\mathtt{Im}[z^3]$.}. In this case $\rho = J_1$. The $\P$-matrix is
$$ \P \ = \ \begin{pmatrix}
4 J_1 & 6 J_2 & 6 J_3 \\
6 J_2 & 9 J_1^2 & 0 \\
6 J_3 & 0 & 9 J_1^2 \end{pmatrix} \ ; $$ and $\P^{(0)}$ is just
the $\P_{11}$ entry, hence
$$ \Q \ = \ 4 \, c_1 \, J_1 \ , \ \ Q \ = \ 4 \, c_1 \ . $$

\bigskip\noindent
{\bf Example 4.} 
Consider $M = \R^3$ and $G = SO(2) \times Z_2$, with $SO(2)$
acting as rotations in the $(x,y)$ plane,  i.e. with the same $R$
as in Example 3, and with $Z_2$ acting as reflections in the $z$
variable,  $z \to - z$ (this is met in studying isotropic nematic
liquid crystal). Here we have two basic invariants,
$$ J_1 \ = \ (x^2 + y^2) \ , \ \ J_2 = z^2 \ ; $$ both of them are
quadratic and they have no algebraic relation. Here again $\rho =
J_1 + J_2$. Now
$$ \P \ = \ \P^{(0)} \ = \ \begin{pmatrix} 4 J_1 & 0 \\ 0 & 4 J_2 \end{pmatrix} $$ and of course
$$ \Q \ = \ \begin{pmatrix} 4 c_1 J_1 \\ 4 c_2 J_2 \end{pmatrix} \ ; \ \
Q \ = \ \begin{pmatrix} 4 c_1 & 0 \\ 0 & 4 c_2 \ . \end{pmatrix} $$
Note that the dimension of the MIB and the expression of $Q$ (and
$\rho$ as well) are the same as in Example 1; thus the two
examples will be dealt with in a similar manner for what concerns
the orbit space analysis, albeit the interpretation in the
original (order parameters) space will be different due to the
different expression for $J_1,J_2$ in terms of the order
parameters.

\bigskip\noindent
{\bf Example 5.} 
Consider now $M = \R^3$ and the group  $G = Z_2 \times Z_2 \times
Z_2 \times S_3$ generated by reflections in each of the
coordinates, i.e. by
$$ \begin{array}{l}
g_x : (x,y,z) \to (-x,y,z), \ \ g_y  :  (x,y,z) \to (x,-y,z), \
g_z : (x,y,z) \to (x,y,-z); \end{array} $$  and by permutations in
the $(x,y,z)$ coordinates. This is the situation studied by
Sergienko, Gufan and Urazhdin \cite{SGU}\footnote{More precisely,
they considered the crystallographic group $Pm3m$ \cite{Ham}; this
includes the reflections in each of the coordinate planes $(xy)$,
$(yz)$, $(xz)$ mentioned above as well as inversion across an axis
($i_x : (x,y,z) \to (x,-y,-z)$ and the like for $i_y$ and $i_z$),
reflections in planes through the diagonal of the unit cube, i.e.
in the planes $y = \pm x$, $z = \pm x$, $z = \pm y$, and rotations
around the coordinate axes. All these elements are also generated
by the $g_x,g_y,g_z$ elements plus the full permutation group in
three elements.}. The MIB consists of three invariants,
$$ J_1 = x^2 + y^2 + z^2 \ , \ \ J_2 = x^2 y^2 + y^2 z^2 + x^2 z^2 \ , \ \ J_3 = x^2 y^2 z^2 \ ; $$
only one of these is quadratic, and $\rho = J_1$. In this case
$$ \P \ = \ 4 \ \begin{pmatrix}J_1 & 2 J_2 & 3 J_3 \\
2 J_2 & (J_1 J_2 + 3 J_3) & 2 J_1 J_3 \\ 3 J_3 & 2 J_1 J_3 & J_2
J_3 \end{pmatrix} \ ; $$ the quadratic part reduces to $\P_{11}$
and is hence scalar, i.e. $\P^{(0)} = 4 J_1$. In this case $\Q = 4
c_1 J_1$, $Q = 4 c_1$.

\bigskip\noindent
{\bf Example 6.} 
Consider now $M = \R^3$ and $G$ the crystallographic group $G =
D_{2h}$ (this is also met in studying anisotropic nematic liquid
crystals); this acts in $\R^3$ via:
\begin{itemize}
\item[(i)] inversion through the center,
$$ I : (x,y,z) \to (-x,-y-z) \ ; $$
\item[(ii)] reflections in each of the coordinate planes:
$$ \begin{array}{l} \s_{xy} : (x,y,z) \to (x,y,-z), \
\s_{yz} : (x,y,z) \to (-x,y,z) , \ \s_{xz} : (x,y,z) \to (x,-y,z)
; \end{array} $$
\item[(iii)] rotations (by an angle $\pi$) around each
coordinate axis,
$$ \begin{array}{l}
R_x : (x,y,z) \to (x,-z,y) , \ R_y : (x,y,z) \to (z,y,-x) , \ R_z
: (x,y,z) \to (-y,x,z) ; \end{array} $$
\item[(iv)] and, of course, the identity.
\end{itemize}

It follows immediately from (i) and (ii) that functions $f(x,y,z)$
can be invariant under $G=D_{2h}$ only if they are actually
functions of $x^2,y^2,z^2$; it is then easy to check that, in view
of (iii) they must actually be also symmetric under any
permutation of the coordinates.

We conclude that the MIB is provided by
$$ J_1 = x^2 + y^2 + z^2, \
J_2 = x^2 y^2 + x^2 z^2 + y^2 z^2 , \
J_3 = x^2 y^2 z^2. $$

This is the same set of invariants met in Example 5 above, and the
discussion for that case (above and in the following) also applies
here; we will thus not further discuss this example. This
illustrates an important point, i.e. that different groups can
give raise to the same set of invariants and can hence be dealt
with, from our point of view, in exactly the same way (see also
the note at the end of Example 4).

\subsection{Straightforward reduction -- non maximal order}
\label{sec:ExII}

We can apply the general reduction scheme discussed earlier on to
the Examples considered above, which we will do now for what
concerns non-maximal order terms. Note that in Examples 1,2 and 4
we would have a polynomial  of order four and no terms of
non-maximal order to simplify. Thus for these examples we will
consider a Landau polynomial of order six, and discuss the
simplification of the terms of order four. In Example 5 and 6, on
the other hand, the lowest possible truncation of the Landau
polynomial is at order twelve, and there is no need to
artificially increase it.

The main purpose of this subsection is to show that one can obtain
very explicit formulas for the generating functions appearing in
our reduction procedure.

For terms of the non-maximal orders the strategy is simple:
simplify them as much as possible. Terms of maximal order will be
dealt with -- taking care of thermodynamic stability -- in the
next subsection.

\medskip\noindent
{\bf Remark 2.} The reader should be warned that we are using a
simplified notation. In fact, each change of variables at order
$m$ will affect the terms of higher orders, hence will change the
coefficients $k_q$ (with $q > m$) appearing in the Landau
potential. In the explicit formulas below for $a_m$, the $k_m$
should be understood as the coefficients appearing in the
potential after all the previous steps have been performed. This
should be taken into account when working concrete applications,
as in the companion paper devoted to application of this method to
liquid crystals \cite{GLC13}.

\bigskip\noindent
{\bf Example 1 (continued).}
In this case the quadratic part of the Landau polynomial will be written as
$$ \Phi_0 \ = \ c_1 \, J_1 \ + \ c_2 \, J_2 \ ; $$
the general invariant polynomial of order four is
$$ \Phi_2 \ = \ k_1 \, J_1^2 \ + \ k_2 \, J_2^2 \ + \ k_3 \, J_1 \, J_2 \ . $$
A similar general expression also holds for the invariant
generating function of order four \footnote{Here and in the other
examples the generating function has a minus sign for convenience
in writing the homological equation and the final results.},
$$ H_2 \ = \  - \ \( a_1 \, J_1^2 \ + \ a_2 \, J_2^2 \ + \ a_3 \, J_1 \, J_2 \) \ . $$

Acting with $\L_0$ on $H_2$, we obtain
\begin{eqnarray*} \L_0 (H_2) &=& [(4 c_1 J_1) \, (2 a_1 J_1 + a_3 J_2) \ + \ (4 c_2 J_2) \, (2 a_2 J_2 + a_3 J_1 ) ] \\
 &=& 8 a_1 c_1 \, J_1^2 \ + \ 8 a_2 c_2 \, J_2^2 \ + \ 4 a_3 (c_1 + c_2) \, J_1 J_2 \ . \end{eqnarray*}
In order to eliminate the term $\Phi_2$, we should solve the
homological equation  $\L_0 (H_2) = \Phi_2$, which is now an
equation for the unknown coefficients $(a_1,a_2,a_3)$ appearing in
the generating function. Actually the homological equation is
promptly recast in terms of the coefficients -- and of the
parameters $c_i$ -- as
$$ 8 a_1 c_1 = k_1 \ , \ \ 8 a_2 c_2 = k_2 \ , \ \ 4 a_3 (c_1 + c_2) = k_3 \ ; $$
thus it suffices to choose
$$ a_1 = \frac{k_1}{8 c_1} \ , \ \ a_2 = \frac{k_2}{8 c_2} \ , \ \
a_3 = \frac{k_3}{4 (c_1 + c_2)} $$ in order to eliminate
completely the $\Phi_2$ term. Needless to say, this is possible
provided $c_1 \not= 0$, $c_2 \not= 0$, $(c_1 + c_2) \not= 0$; if
some of these non-degeneracy conditions (involving only the
quadratic part of the Landau polynomial) fails, we correspondingly
have to retain the associated quartic term(s).

\bigskip\noindent
{\bf Example 2 (continued).}
In this case the quadratic part of the Landau polynomial will be written as
$$ \Phi_0 \ = \ c_1 \, J_1 \ + \ c_2 \, J_2 \ + \ c_3 \, J_3 \ ; $$
the general invariant polynomial of order four is
$$ \Phi_2 \ = \ k_1 \, J_1^2 \ + \ k_2 \, J_2^2 \ + \ k_3 \, J_3^2 \ + \
k_4 \, J_1 J_3 \ + \ k_5 \, J_2 J_3  $$
(note we have not written any $J_1 J_2$ term; as $J_1 J_2 =
J_3^2$, this would be redundant). A similar general expression
also holds for the invariant generating function of order four,
$$ H_2 \ = \  - \ \( a_1 \, J_1^2 \ + \ a_2 \, J_2^2 \ + \ a_3 \, J_3^2 \ + \
a_4 \, J_1 J_3 \ + \ a_5 \, J_2 J_3 \) \ . $$

Acting with $\L_0$ on $H_2$, we obtain
\begin{eqnarray*} \L_0 (H_2) &=& 4 a_1 c_1 \, J_1^2 \ + \ 4 a_2 c_2 \, J_2^2 \ + \
[ (a_4 + a_5) c_3 + 2 a_3 (c_1 + c_2 + 2 c_3)] \, J_3^2 \\
& & \ + \ [2 a_1 c_3 + a_4 (3 c_1 + c_2 + 2 c_3)] \, J_1 J_3 \ + \
[2 a_2 c_3 + a_5 (c_1 + 3 c_2 + 2 c_3)] \, J_2 J_3  \ . \end{eqnarray*}
The homological equation $\L_0 (H_2) = \Phi_2$ is now recast as
\begin{eqnarray*}
k_1 &=& 4 a_1 c_1 \ , \ \ k_2 \ = \  4 a_2 c_2 \ , \ \ k_3 \ = \
(a_4 + a_5) c_3 + 2 a_3 (c_1 + c_2 + 2 c_3) \ , \\ k_4 &=& 2 a_1
c_3 + a_4 (3 c_1 + c_2 + 2 c_3) \ , \ \ k_5 \ = \ 2 a_2 c_3 + a_5
(c_1 + 3 c_2 + 2 c_3)]  \ . \end{eqnarray*} Its solution is
provided by
\begin{eqnarray*} a_1 &=& \frac{k_1}{4 c_1} , \ a_2 = \frac{k_2}{4 c_2} , \ \
a_3 = \left[ 2 k_3 \ + \ \frac{c_3 (c_3 k_1 - 2 c_1 k_4)}{c_1 (3
c_1 + c_2 + 2 c_3)} \ + \ \frac{c_3 (c_3 k_2 - 2 c_2 k_5)}{c_2
(c_1 + 3 c_2 + 2 c_3)}\right] \,
\left[ \frac{1}{4 \, (c_1 + c_2 + 2 c_3) } \right] , \\
a_4 &=& \frac{2 c_1 k_4 - c_3 k_1}{2 c_1 (3 c_1 + c_2 + 2 c_3)} ,
\ a_5 = \frac{2 c_2 k_5 - c_3 k_2}{2 c_2 (c_1 + 3 c_2 + 2 c_3)} \
. \end{eqnarray*} Here again one should impose non-degeneracy
conditions corresponding to the requirement that the fractions
appearing in the explicit expressions for the $a_i$ are well
defined; if these fail, some (or all) of the fourth-order terms
cannot be eliminated.

\bigskip\noindent
{\bf Example 3 (continued).} The situation in Example 3 is
different from the one of previous examples; indeed, now we have
third order invariants. We write as usual
$$ \Phi_0 \ = \ c_1 \ J_1^2 \ , $$
and the third order invariant term will be
$$ \Phi_1 \ = \ k_1 \, J_2 \ + \ k_2 J_3 \ ; $$
correspondingly the third order generating functions will be written as
$$ H_1 \ = \  - \ \( a_1 \, J_2 \ + \ a_2 \, J_3 \) \ , $$
and does not depend on $J_1$. Thus any third order invariant is in
the kernel of $\L_0$, and it does not produce any effect (at the
first order level). At order four, the invariant term and the
generating functions will be written as
$$ \Phi_2 \ = \ k_3 \, J_1^2 \ , \ \ H_2 \ = \  - \, a_3 \, J_1^2 \ . $$
We immediately have
$$ \L_0 (H_2) \ = \ 8 \, c_1 \, a_3 \ J_1^2 \ , $$
and the homological equation is solved by choosing
$$ a_3 \ = \ \frac{k_3}{8 c_1} \ . $$

We also have invariants (and generating functions) of order five;
these are
$$
\Phi_3 \ = \ k_4 \, J_1 J_2 \ + \ k_5 \, J_1 J_3 \ ; \ \ H_3 \ = \
 - \ \( a_4 \, J_1 J_2 \ + \ a_5 \, J_1 J_3 \) \ . $$ Now we
have
$$ \L_0 (H_3) \ = \ 4 c_1 a_4 \ J_1 J_2 \ + \ 4 c_1 a_5 \ J_1 J_3 \ , $$
hence in order to solve the homological equation we just choose
$$ a_4 \ = \ \frac{k_4}{4 c_1} \ , \ \ a_5 \ = \ \frac{k_5}{4 c_1}\ . $$
Needless to say, the elimination of terms of both order four and
five is possible only under the condition $c_1 \not= 0$.

\bigskip\noindent
{\bf Example 4 (continued).} As remarked above, once we set our
problem in orbit space this is the same as Example 1 (albeit with
a different interpretation when we want to go back to the order
parameters space). We can thus just reproduce the computations
seen in dealing with Example 1 above. We write the quadratic term
in the form
$$ \Phi_0 \ = \ c_1 \, J_1 \ + \ c_2 \, J_2 \ ; $$
the quartic term and generating functions will read
\begin{eqnarray*} \Phi_2 &=& k_1 J_1^2 \ + \ k_2 J_2^2 \ + \ k_3 J_1 J_2 \ ; \ \
H_2 \ = \  - \ \( a_1 J_1^2 \ + \ a_2 J_2^2 \ + \ a_3 J_1 J_2 \) \
.
\end{eqnarray*} Thus we readily get
$$ \L_0 (H_2) \ = \ 8 a_1 c_1 \, J_1^2 \ + \ 8 a_2 c_2 \, J_2^2 \ + \ 4 a_3 (c_1 + c_2) J_1 J_2 \ ; $$
the homological equation is solved by choosing
$$ a_1 = \frac{k_1}{8 c_1} \ , \ \  a_2 = \frac{k_2}{8 c_2} \ , \ \ a_3 = \frac{k_3}{4 (c_1 + c_2)} \ . $$
In this case the non-degeneracy conditions are of course $c_1 \not=0$, $c_2 \not= 0$, $(c_1+c_2) \not= 0$.

\bigskip\noindent
{\bf Examples 5 \& 6 (continued).} We will as usual write $\Phi_0
= c_1 J_1$. In this case no terms of odd order are  allowed by the
symmetry, and we should eliminate as many terms as possible at
orders 4,6,8,10.

The different invariant terms of order not higher than ten are written as
{\small
\begin{eqnarray*}
\Phi_2 &=& k_1 J_1^2 + k_2 J_2 \ , \\
\Phi_4 &=& k_3 J_1^3 + k_4 J_1 J_2 + k_5 J_3 \ , \\
\Phi_6 &=& k_6 J_1^4 + k_7 J_1^2 J_2 + k_8 J_1 J_3 + k_9 J_2^2 \ , \\
\Phi_8 &=& k_{10} J_1^5 + k_{11} J_1^3 J_2 + k_{12} J_1^2 J_3 +
k_{13} J_1 J_2^2 + k_{14} J_2 J_3 \ ; \end{eqnarray*} } the
expressions for the $H_k$ are obtained from these by replacing the
$k_i$ coefficients with $a_i$ ones and changing sign. As we have
seen above, $$ \Q \ = \ 4 \ c_1 \ J_1 \ , \ \ Q \ = \ 4 \ c_1 \ .
$$ This means that, as easy to compute,
\begin{eqnarray*}
\L_0 (H_2) &=& 8 a_1 c_1 J_1^2 \ , \\
\L_0 (H_4) &=& 12 a_3 c_1 J_1^3 \ + \ 4 a_4 c_1 J_1 J_2 \ , \\
\L_0 (H_6) &=& 16 a_6 c_1 J_1^4 \ + \ 8 a_7 c_1 J_1^2 J_2 + 4 a_8 c_1 J_1 J_3 \ , \\
\L_0 (H_8) &=& 20 a_{10} c_1 J_1^5 + 12 a_{11} c_1 J_1^3 J_2 +
8 a_{12} c_1 J_1^2 J_3 + 4 a_{13} c_1 J_1 J_2^2  \ . \end{eqnarray*}

Note that in this case several terms appearing in $\Phi$ are {\it
not } in the range of $\L_0$, and hence cannot be eliminated. At
the same time, some terms in the $H_k$ are in the kernel of $\L_0$
and thus inessential to our procedure.

In particular, by choosing
\begin{eqnarray*}
a_1 &=& \frac{k_1}{8 c_1} \ , \ \
a_3 \ = \ \frac{k_3}{12 c_1} \ , \ \
a_4 \ = \ \frac{k_4}{4 c_1} \ , \ \
a_6 \ = \ \frac{k_6}{16 c_1} \ , \ \
a_7 \ = \ \frac{k_7}{8 c_1} \ , \\
a_8 &=& \frac{k_8}{4 c_1} \ , \ \ a_{10} \ = \ \frac{k_{10}}{20
c_1} \ , \ \ a_{11} \ = \ \frac{k_{11}}{12 c_1} \ , \ \ a_{12} \ =
\ \frac{k_{12}}{8 c_1} \ , \ \ a_{13} \ = \ \frac{k_{13}}{4 c_1} \
, \end{eqnarray*} the Landau potential is reduced to one of the
form
$$ \^\Phi \ = \ \b_1 J_2 \ + \ \b_2 J_3 \ + \ \b_3 J_2^2 \ + \ \b_4 J_2 J_3 \ . $$
Needless to say, this is possible provided $c_1 \not= 0$. It would
be possible (but not of interest here) to compute explicitly the
coefficients $\b_i$ in terms of the $k_i$ and $a_i$. Note that the
coefficients $a_2, a_5, a_9, a_{14}$ are not determined by our
procedure.

\subsection{Straightforward reduction -- maximal order}
\label{sec:ExIII}

For terms of maximal order our strategy should {\it not} be to
eliminate whatever can be eliminated through our procedure. We
should instead take care to guarantee the thermodynamic stability
(i.e., convexity for large $|x|$, see section \ref{sec:stability})
of the simplified (truncated) Landau polynomial.

A simple way to guarantee this, if all terms could be eliminated,
is by arranging things so that $\Phi_N = |x|^{N+2}$. (Note this is
surely possible: all terms which can be eliminated are in the
range of $\L_0$, so we can also arrange things so that a specific
term appears as result of applying $\L_0$ on a suitable generating
function.) In practice, we should express $\rho = |x|^2$ in terms
of the basic polynomials $J_i$, and then keep the term $\rho^m$
(with $m = N/2 + 1$) in $\Phi_N$. \footnote{If some term in
$\Phi_N$ is resonant and cannot be eliminated, then we should
enter into details of the term, and see how we can guarantee
convexity for large $|x|$.}

\bigskip\noindent
{\bf Example 1 (continued).} In this case $N=4$, and
$$ \Psi_4 \ = \ k_4 J_1^3 + k_5 J_1^2 J_2 + k_6 J_1 J_2^2 + k_7 J_2^3 \ ; $$
the general generating function at this order can be written as
$$ \chi_4 \ = \ b_1 J_1^3 + b_2 J_1^2 J_2 + b_3 J_1 J_2^2 + b_4 J_2^3
\ . $$ Acting on this with $\L_0$, we obtain
\begin{eqnarray*} \L_0 (\chi_4) &=& 12 b_1 c_1 J_1^3 +
4 J_2 \[ b_2 (2 c_1 + c_2) J_1^2  + J_2 (b_3 (c_1 + 2 c_2) J_1 + 3
b_4 c_2 J_2) \] \ . \end{eqnarray*} Recalling that the transformed
term will be
$$ \wt{\Psi_4} \ = \ \Psi_4 + \de \Psi_4 \ = \ \Psi_4 \ - \ \L_0 ( \chi_4 )  \ , $$
and that in this case $\rho = J_1 + J_2$, it is possible to get $
\wt{\Psi_4} = \b \rho^3$ by choosing (provided $c_2\not= 0$, $c_1
+ c_2 \not= 0$)
\begin{eqnarray*} b_2 &=& \frac{36 b_1 c_1 - 3 k_4 + k_5}{4 (2 c_1 + c_2)} \ , \ \
b_3 \ = \ \frac{36 b_1 c_1 - 3 k_4 + k_6}{4 (2 c_1 + c_2)} \ , \ \
b_4 \ = \ \frac{12 b_1 c_1 - k_4 + k_7}{12 c_2} \ . \end{eqnarray*}
With this choice, we have
$$ \wt{\Psi_4} \ = \ (k_4 - 12 b_1 c_1 ) \ \rho^3 \ ; $$
thus, by choosing (as usual, under the assumption $c_1 \not= 0$)
$$ b_1 \ = \ \frac{k_4 - 1}{12 \, c_1} $$ we always obtain
$\wt{\Psi_4} = \rho^3$.

Summarizing, in this case ($G = Z_2 \times Z_2$, Landau polynomial
of order six), under the assumptions $c_1 \not= 0$, $c_2 \not= 0$,
$(c_1 + c_2 ) \not= 0$  we can always reduce to consider a Landau
polynomial of the form
$$ \Psi \ = \ c_1 \, J_1 \ + \ c_2 \, J_2 \ + \ (J_1 + J_2)^3 \ , $$
thus getting rid of the seven additional parameters $k_1,...,k_7$.

\bigskip\noindent
{\bf Example 1B.} In Example 1 we have considered the case where
one considers a Landau polynomial of order six (this was in order
to avoid a trivial case at non-maximal orders, i.e. in
Sect.\ref{sec:ExII}). We can now also consider the case where the
Landau polynomial is of order four (thus $N=2$), i.e. $\Phi =
\Phi_0 + \Phi_2$. The quadratic part $\Phi_0$ is not modified, so
we only have to modify the quartic -- and maximal -- term
$\Phi_2$. We know from Example 1 that it could be fully
eliminated, but in order to preserve stability we should actually
leave a term of the type $\rho^2$, i.e. $(J_1 + J_2 )^2$. In view
of the expressions for $\Psi_2$ and $\chi_2$ (see again Example 1
in Sect.\ref{sec:ExII}), this is obtained by choosing
$$ a_1 = \frac{k_1 -1}{8 c_1} , \ a_2 = \frac{k_2 - 2 }{4 (c_1 + c_2)} , \
a_3 = \frac{k_3- 1}{8 c_2} \ . $$ Thus for $G = Z_2 \times Z_2$,
and Landau polynomial of order four, under the assumption $c_1
\not= 0$, $c_2 \not= 0$, $(c_1 + c_2 ) \not= 0$  we can always
reduce to consider a Landau polynomial of the form
$$ \Psi \ = \ c_1 \, J_1 \ + \ c_2 \, J_2 \ + \ (J_1 + J_2)^2 \ , $$
thus getting rid of the three additional parameters $\{k_1,k_2,k_3\}$.

\bigskip\noindent
{\bf Example 2 (continued).} In this case also $N=4$, but we
should take into account that $J_1 J_2 = J_3^2$; we can thus write
the general $\Psi_4$ term and the generating function $\chi_4$ as
\begin{eqnarray*} \Psi_4 &=& k_6 J_1^3 + k_7 J_1^2 J_3 + k_8 J_1 J_3^2 +
k_9 J_2^2 J_3 + k_{10} J_2 J_3^2  + k_{11} J_2^3 + k_{12} J_3^3 \ ; \\
\chi_4 &=& b_1 J_1^3 + b_2 J_1^2 J_3 + b_3 J_1 J_3^2 + b_4 J_2^2 J_3 +
b_5 J_2 J_3^2 + b_6 J_2^3 + b_7 J_3^3 \ . \end{eqnarray*}

The explicit expressions (which can be readily obtained with a
symbolic manipulation language -- e.g. in Mathematica) are in this
case rather involved and we will not report them. However, one
obtains that by a suitable choice of the coefficients appearing in
$\chi_4$, it is possible to obtain
$$ \wt{\Psi_4} \ := \ \Psi_4 \ - \ \L_0 (\chi_4) \ = \ \rho^3 \ = \ (J_1 + J_2)^3  \ . $$
The suitable choices for the $b_i$ have a denominator $d_i$ of the
form $d_i =  r_i \ (c_1 + c_2) \ R $ where  $r_i$ are some
positive integers and \begin{eqnarray*} R &=& (5 c_1^2 + 26 c_1
c_2 + 5 c_2^2 - 4 c_3^2) \ (4 c_1 c_2 - c_3^2) \  (8 c_1^2 + 20
c_1 c_2 + 8 c_2^2 - c_3^2) \ . \end{eqnarray*} The non-degeneracy
conditions allowing for such a reduction of the Landau polynomial
(beside those met at order four, see Sect.\ref{sec:ExII}) are then
just the non-vanishing of the above denominators, i.e. $(c_1 +
c_2) \not= 0$, $R \not= 0$.

Summarizing, in this case ($G = Z_2$, Landau polynomial of order
six), under the assumption $c_1 \not= 0$, $c_2 \not= 0$, $(c_1 +
c_2) \not= 0$, $(c_1 + c_2 + 2 c_3) \not= 0$, $(3 c_1 + c_2 + 2
c_3) \not= 0$, $(c_1 + 3 c_2 + 2 c_3) \not= 0$, $R \not= 0$, we
can always reduce to consider a Landau polynomial of the form
$$ \Psi \ = \ c_1 \, J_1 \ + \ c_2 \, J_2 \ + \ c_3 \, J_3 \ + \ (J_1 + J_2)^3 \ , $$
thus getting rid of the twelve additional parameters $k_1,...,k_{12}$.

\bigskip\noindent
{\bf Example 2B.} In this case as well we can consider a variant
of the above example, namely the case where the Landau polynomial
is of order four ($N=2$), $\Phi = \Phi_0 +\Phi_2$ and hence the
discussion in the framework of Sect.\ref{sec:ExII} would have been
trivial.

In this case
\begin{eqnarray*}
\Psi_2 &=& k_1 \, J_1^2 \ + \ k_2 \, J_1 J_3 \ + \ k_3 \, J_2 J_3 \ + \ k_4 \, J_2^2 \ + \ k_5 J_3^2 \ ; \\
\chi_2 &=& a_1 \, J_1^2 \ + \ a_2 \, J_1 J_3 \ + \ a_3 \, J_2 J_3
\ + \ a_4 \, J_2^2 \ + \ a_5 J_3^2 \ . \end{eqnarray*} Explicit
formulas are still rather involved; in fact now $\wt{\Psi}_2 =
\Psi_2 - \L_0 (\chi_2)$ results to be simply $\wt{\Psi}_2 =
\rho^2$ with the choice
$$ a_i \ = \ \frac{\alpha_i}{ r_i \ A} $$
where $\alpha_i$ are some rather involved polynomial in the $c_i$
and $k_i$, the  $r_i$ are positive integers, and
\begin{eqnarray*}
A &=& (c_1 + c_2) \ [12 c_1^3 c_2 - 3 c_2^2 c_3^2 + c_3^4  +
  c_1^2 (40 c_2^2 - 3 c_3^2) + 2 c_1 (6 c_2^3 - 7 c_2 c_3^2)] \ . \end{eqnarray*}
This allows to explicitly identify the non-degeneracy conditions under which such a reduction is possible.

\bigskip\noindent
{\bf Example 3 (continued).} In this case the general term
$\Psi_4$ and the generating function $\chi_4$ of order six read
\begin{eqnarray*}
\Psi_4 &=&  k_6 \, J_1^3 \ + \ k_7 \, J_2^2 \ + \ k_8 \, J_3^2 \ + \ k_9 \, J_2 \, J_3 \ ; \\
\chi_4 &=&  b_1 \, J_1^3 \ + \ b_2 \, J_2^2 \ + \ b_3 \, J_3^2 \ + \ b_4 \, J_2 \, J_3 \ . \end{eqnarray*}
We have
$$ \L_0 ( \chi_4 ) \ = \ 12 \ b_1 \ J_1^3 \ , $$
thus the terms other than $J_1^3$ cannot be eliminated. On the
other hand, in this case $\rho = J_1$, so this is precisely the
term which we do not want to cancel (the reader can easily check
that setting this to zero produces direction in which the
potential is not convex for large $|\xb |$); we can however set
this to unity, which is obtained by setting
$$ b_1 \ = \ \frac{k_6 - 1}{12} \ . $$

Summarizing, in this case -- i.e. for $G = {\bf Z}_3$ -- the sixth
order Landau polynomial can always (provided $c_1 \not= 0$) be
reduced to
$$ \Phi \ = \ c_1  J_1 \, + \, k_1  J_2 \, + \, k_2  J_3 \, + \, k_7  J_2^2 \, + \,
k_8  J_3^2 \, + \, k_9  J_2  J_3 \, + \, J_1^3 $$ (recall
$k_7,k_8,k_9$ are in general different from the initial ones),
thus getting rid of the four additional parameters $\{ k_3 , ... ,
k_6 \}$.

\bigskip\noindent
{\bf Example 4 (continued).} In orbit space this is the same as
Example 1; the computations and results would just reproduce those
seen in dealing with Example 1 above, and are thus omitted.

\bigskip\noindent
{\bf Example 5 \& 6 (continued).} In this case the general term
$\Psi_{10}$ and the generating function $\chi_{10}$ of order
twelve are written as
\begin{eqnarray*}
\Psi_{10} &=& k_{15} J_1^6 + k_{16} J_1^4 J_2 + k_{17} J_1^3 J_3 + k_{18} J_1^2 J_2^2  +
k_{19} J_1 J_2 J_3 + k_{20} J_2^3 + k_{21} J_3^2 \ ; \\
\chi_{10} &=& b_{1} J_1^6 + b_{2} J_1^4 J_2 + b_{3} J_1^3 J_3 + b_{4} J_1^2 J_2^2  +
b_{5} J_1 J_2 J_3 + b_{6} J_2^3 + b_{7} J_3^2 \ . \end{eqnarray*}
We obtain immediately
\begin{eqnarray*} \L_0 (\chi_{10} ) &=& (k_{15} - 24 b_1 c_1) \, J_1^6 \ + \ (k_{16} - 16 b_2 c_1) \,
J_1^4 \, J_2 \ + \ (k_{17} - 12 b_3 c_1) \, J_1^3 \, J_3 \\
 & & + \ (k_{18} - 8 b_4 c_1) \,  J_1^2 \, J_2^2 \ + \ (k_{19} - 4
b_5 c_1) \, J_1 \, J_2 \, J_3 \ + \ k_{20} \, J_2^3 \ + \ k_{21}
\, J_3^2 \ . \end{eqnarray*} It is thus clear that we could cancel
the $J_1^6$ term (which we do not actually want to cancel) and we
can cancel all the other terms at the exception of the $J_2^3$ and
the $J_3^2$ ones, just by choosing, under the assumption $c_1
\not= 0$,
$$ b_1 = \frac{k_{15}}{24 c_1} , \ b_2 = \frac{k_{16}}{16 c_1} , \
b_3 = \frac{k_{17}}{12 c_1} , \ b_4 = \frac{k_{18}}{8 c_1} , \ b_5
= \frac{k_{19}}{4 c_1} \ . $$ On the other hand, the coefficients
$b_6$ and $b_7$ are inessential (the corresponding terms are in
the kernel of $\L_0$). As for $b_1$, with the choice
$$ b_1 \ = \ \frac{k_{15} - 1}{24 c_1} $$
we will have a term $\rho^6$ in $\wt{\Psi}_6$.

In fact, with these choices (and those considered in
Sect.\ref{sec:ExII} for lower order generating functions, all of
them valid under $c_1 \not= 0$), the Landau polynomial of order
twelve is reduced to
\begin{eqnarray*}
\^\Phi &=& c_1 J_1 \ + \ \b_1 J_2 \ + \ \b_2 J_3 \ + \ \b_3 J_2^2 \ + \ \b_4 J_2 J_3 \ + \
\b_5 J_2^3 \ + \ \b_6 J_3^2 \ + \ J_1^6 \ ; \end{eqnarray*}
this depends on 6 parameters, while the original one depended on 22 parameters.

Note also that the convexity for large $|{\bf x}|$ is guaranteed,
precisely by the presence of the $J_1^6$ term.

\section{Adapted coordinates}
\label{sec:adapt}

Let us go back to considering \eqref{eq:L0K}; in that formula $
Q_\a^\b = c_\ga K_{\ga \a}^\b$ is by construction a numerical
matrix. It is quite clear that we would be better off using a set
of quadratic invariants such that the matrix $Q$ characterizing
the homological operator $\L_0$ had diagonal form; if this is not
possible, one could at least set $Q$ in Jordan normal form.

Let us denote a set of new quadratic invariants as \beql{eq:Z}
Z_\mu \ = \ A_{\mu \nu} \, J_\nu \ ; \eeq here $A$ is a constant
matrix; correspondingly we have
$$ J_\a \ = \ A^{-1}_{\a \b} \, Z_\b \ \ , \ \ \
(\pa / \pa J_\a) \ = \ A^T_{\a \b} \, (\pa / \pa Z_\b) \ . $$ The
operator $\L_0$ defined in \eqref{eq:L0K} reads, with these basis
invariants, \beql{eq_L0QD} \L_0 \ = \ \[ \( A_{\mu \a} \, Q_{\a
\b} \, A^{-1}_{\b \nu} \) \, Z_\nu \] \ \frac{\pa}{\pa Z_\mu} \ :=
\ (P_{\mu \nu} \, Z_\nu ) \ \frac{\pa}{\pa Z_\mu} \ ; \eeq here
the matrix $P$ is given by \beql{eq:P} P \ = \ A \, Q \, A^{-1} \
. \eeq We also write $P = P_s + P_n$,  with  $P_s$ and $P_n$ the
semisimple and nilpotent parts \footnote{The nilpotent part could
(and will most often) vanish, but we are not guaranteed this will
be the case in general.} of $P$, with \beql{eq:Ps} P_s \ = \
\mathrm{diag} (\la_1 , ... , \la_s ) \ . \eeq  Needless to say, to
reach this form the matrix $A$ in \eqref{eq:Z} should be chosen
precisely as the matrix taking $Q$ into Jordan normal form, which
we assume below.

In the following, we will use the set $Z_\a$ ($\a = 1,...,s$) of
quadratic invariants, and write $\z_i = J_{s+i}$ ($i=1,...,q=r-s$;
if $s=r$ then no $\z$ is present) for basic invariants of higher
order.

\subsection{Semisimple $P$}

Let us consider the case where $P_n = 0$. In this case we can
consider the monomials (in the invariants) \beql{eq:Ga} \Ga_{{\bf
k} {\bf h}} \ := \ Z_1^{k_1} ... Z_s^{k_s} \, \z_1^{h_1} ...
\z_q^{h_q} \ ; \eeq note that $\chi_m$ can be written as
\beql{eq:chiga} \chi_m \ = \ \ga_{k_1 , ... , k_s ; h_1,...,h_q}
\, \Ga_{{\bf k} {\bf h}} \ , \eeq where the sum extends on all the
sets ${\bf k},{\bf h}$ such that \beql{eq:condm} \sum_{\a=1}^s 2
k_\a \ + \ \sum_{i=1}^q d_{r+i} \, h_i \ = \ m \, + \, 2 \ . \eeq
It follows immediately from \eqref{eq:P} and \eqref{eq:Ps} that
\beql{eq:L0act} \L_0 \( \Ga_{{\bf k} {\bf h}} \) \ = \ (\la_\a
\cdot k_\a ) \ \Ga_{{\bf k} {\bf h}} \ . \eeq Thus, the kernel of
$\L_0$ restricted to $S_m$ is spanned by all the $\Ga$ -- among
those satisfying \eqref{eq:condm} -- such that \beql{eq:res}
\sum_{\a = 1}^s \la_\a \cdot k_\a \ = \ 0 \ . \eeq This is the
equivalent of the Poincar\'e resonance condition in our case; thus
we will call terms $\Ga_{{\bf k} {\bf h}}$ satisfying it, {\it
resonant}, and \eqref{eq:res} will be said to be the {\it
resonance condition}.

Similarly, the range of $\L_0$ (applied to $S_m$) is the subspace
of $S_m$ spanned by the $\Ga_{{\bf k} {\bf h}}$ -- among those
satisfying \eqref{eq:condm} -- which do {\it not } satisfy
\eqref{eq:res}.

We conclude that in this case one can always eliminate -- as
usual, by a careful choice of the generating functions
$H_1,H_2,...$, see below -- all terms of higher order which are
not resonant \footnote{Note again here we should not attempt to
eliminate the maximal order terms as these are needed to guarantee
the thermodynamic stability.}. In other words, we can always
reduce -- at least in principles -- to consider Landau polynomials
in which {\it the terms of higher (but not maximal) order which
are allowed by the symmetry but are non-resonant, are absent}.

In more detail, we can always write the generating function as
$H_m =  - \xi_{{\bf k} {\bf h}} \Ga_{{\bf k} {\bf h}}$; if the
term of order $m+2$ is written as $\Phi_m = c_{{\bf k} {\bf h}}
\Ga_{{\bf k} {\bf h}}$, then the homological equation is solved by
choosing \beql{eq:homad} \xi_{{\bf k} {\bf h}} \ = \ \frac{c_{{\bf
k} {\bf h}} }{(\la_\a \cdot k_\a )} \ , \eeq for the ${\bf k}$
satisfying $(\la_\a \cdot k_\a ) \not= 0$, while $\xi_{{\bf k}
{\bf h}}$ is undetermined (we can e.g. set it to zero) for ${\bf
k}$ satisfying \eqref{eq:res}.

Note that our procedure produces hence some (possibly small)
denominators; this will make that the procedure is well defined
only in a small neighborhood of the origin in the $M$ space.
Physically, this is not a problem provided this neighborhood is
large enough to include the symmetry-breaking minima of the
theory; if this is not the case, the procedure described here is
only formal and not helpful in practice. This problem is well
known in the applications of Poincar\'e-Birkhoff normal forms in
dynamical and Hamiltonian systems (see also Sect.\ref{sec:disc}).

\subsection{Non semisimple $P$}

In the case where $P$ is not semisimple, i.e. $P_n \not= 0$, one
does actually proceeds in the same way, dealing with $P_s$ rather
than the full $P$. That is, the system is set in normal form with
respect to the semisimple part of $P$, and resonant terms are
defined with reference to the semisimple part of $P$ alone (i.e.
as above); this is completely analogous to what is done in the
Poincar\'e approach to dynamical systems \cite{ArnG,Elp,CGs}.

As a result, the Landau polynomial can be reduced to include only
resonant higher order terms $\Phi_k$, while the quadratic one is
$\Phi_0 = \Phi_s + \Phi_n$, where of course $\Phi_s, \Phi_n$ are
associated respectively to $P_s$ and $P_n$.

In terms of the operator $\L_0$, this amounts to saying that our
previous results (nonlinear terms can be reduced to those not in
the range of $\L_0$, etc.) remain true, with a difference: now the
operator $\L_0$ is not associated to the full quadratic part
$\Phi_0$, but instead to its semisimple part $\Phi_s$ only.

\subsection{Example}

By looking at Section \ref{sec:ExI}, one  sees that in Example 1
the $Q$ matrix is diagonal, and in Examples 3,4,5 and 6 it
actually reduces to a scalar. Thus the only example, among those
considered above, in which the coordinates are not already adapted
is Example 2; we are now going to consider this.

Note that the matrix $Q$ for this case (see Section \ref{sec:ExI})
is diagonal when $c_3 = 0$; we will thus assume $c_3 \not= 0$, and
simplify our notation by setting $k_1 = c_1/c_3$, $k_2 = c_2 /
c_3$, $\ga = k_1+k_2$. Then $Q$ reads
$$ Q \ = \ c_3 \ \begin{pmatrix} 4 k_1 & 0 & 2 \\
0 & 4 k_2 & 2 \\ 1 & 1 & 2 (k_1 + k_2) \end{pmatrix} \ . $$

This is taken into Jordan normal form by the map
$$ A \ = \ \frac{1}{4 \b^2} \begin{pmatrix} - 2 \de & 2 \de  & 4 \de^2 \\
\de - \b & - (\de + \b) & 2  \\
\de + \b  & - (\de - \b)  & 2 \end{pmatrix} \ , $$
where we have defined
$$ \de := k_1 - k_2 \ ; \ \ \b = \sqrt{1 + \de^2} \ . $$
The inverse matrix is given by
$$ A^{-1} \ = \ \begin{pmatrix} - \de^{-1} & \ (1 + 2 \b \de - 2 \b^2) (\b - \de)^{-1} \ &
\ (-1 + 2 \b \de + 2 \b^2) (\b + \de)^{-1} \ \\ \de^{-1} & - (\b -
\de)^{-1} & (\b + \de)^{-1} \\ 1 & 1 & 1
\end{pmatrix} \ ; $$ the corresponding (diagonal) Jordan form for
$Q$ is
$$ \wt{P} \ = \ 2 \, c_3 \ \begin{pmatrix} \ga & 0 & 0 \\
 0 & \ga - \b & 0 \\ 0 & 0 & \ga + \b \end{pmatrix} \ . $$

In order to check that our construction is working correctly, one
can reach the same result in a different way. The new invariants
$Z_\a = A_{\a \b} J_\b$ are given by
\begin{eqnarray*}
Z_1 &=& [\de/(2 \b^2)] \ (- x^2 + 2 \de x y + y^2 ) \ , \\
Z_2 &=& [(\de - \b)/(4 \b^2)] \ [x^2 - 2 (\de + \b) x y +  (\de + \b)^2 y^2 ] \ , \\
Z_3 &=& [(\de + \b)/(4 \b^2)] \ [x^2 + 2 (\b - \de) x y + (\b -
\de)^2 y^2 ] \ . \end{eqnarray*} From these expression one easily
computes the $\P$-matrix in terms of the $Z$, which we denote by
$\wt{\P}$. On the other hand, we should now express the quadratic
part of the Landau polynomial in terms of the $Z_\a$, i.e. write
$$ \Phi_0 \ = \ c_\a \, J_\a \ = \ c_\a \, A^{-1}_{\a \b} \, Z_\b
\ := \ \wt{c}_\b \, Z_\b . $$ Finally we can write $\L_0 = -
\wt{c}_\a \wt{\P}_{\a \b} (\pa / \pa Z_\b)$; doing this
explicitly, with standard (and boring) algebra we obtain $\L_0 =
(\wt{P}_{\a \b} Z_\b) (\pa / \pa Z_\a)$ with the same $\wt{P}$
given above.

\section{Reduction and analysis of Landau potentials}
\label{sec:reduction}

In this section we will discuss how the reduction studied here can
be used to analyze -- both quantitatively and qualitatively -- the
behavior of concrete physical systems in the framework of Landau
theory, i.e. the critical points of Landau potentials. We will
again refer to the examples considered in the previous section.

\subsection{Quantitative analysis of Landau polynomials}
\label{sec:redA}

\def\xt{\widetilde{x}}
\def\yt{\widetilde{y}}

In this subsection we consider the simplest of the Examples
presented above -- i.e. Example 1 (recall all computations will
also immediately apply to Example 4), albeit with a sixth order
Landau polynomial -- and show how the method depicted here can be
concretely used to study the problem. We will give a complete --
qualitative and quantitative -- analysis of this simple problem.

In concrete cases one would be satisfied in discussing the
qualitative behavior, as will be done in the next subsection, and
the simple case at hand here is just to be meant as an
illustration of the method.

The analysis of a concrete problem requires to obtain definite
expressions for the coefficients appearing in the generating
functions, i.e. to describe exactly the change of variables to be
considered. A discussion of how to obtain these in a
computationally efficient way is contained in the companion paper
\cite{GLC13}, and here we will just provide the resulting formulas
for the normalizing change of coordinates, see below.

\bigskip\noindent
{\bf Example 1 \& 4 (continued).} The general sixth order
$G$-invariant Landau polynomial for the $G$ action considered in
Example 1 (and also applying to Example 4) is given, in terms of
the original $(x,y)$ coordinates, by
\begin{eqnarray*} \Phi &=& (c_1 x^2 + c_2 y^2 ) \ + \ (k_1 x^4 +
k_2 x^2 y^2 + k_3 y^4)  \ + \ (k_4 x^6 + k_5 x^4 y^2 + k_6 x^2 y^4
 + k_7 y^6) \ ; \end{eqnarray*} note this depends on eight
parameters, and analyzing its behavior in terms of these
parameters would be quite a substantial task.

By the (non unique, see above) change of coordinates
\begin{eqnarray*} x \to \xt &=& x \ \[ 1 \ - \ \( \frac{(k_1 x^2 +
k_2 y^2)}{2 c_1} \)  \ + \ \( \frac{(1-k_4)
x^4 + (3 - k_6) y^4}{2 c_1}  + \frac{7 k_1^2 x^4 + 3 k_2^2
y^4}{8 c_1^2}  + \frac{k_2 k_3 y^4}{2 c_1 c_2} \) \] \ , \\
y \to \yt &=& y \ \[ 1 \ - \ \( \frac{k_3 y^2}{2 c_2} \)  \
\( \frac{(3-k_5) x^4 + (1 - k_7) y^4}{2 c_2}
 +
\frac{5 k_1 k_2 x^4}{4 c_1 c_2} + \frac{7 k_3^2 y^4}{8 c_2^2} \)
\] \ , \end{eqnarray*}  and truncating the resulting polynomial
again at order six, the Landau potential is transformed into
$$ \wt{\Phi} \ = \ c_1 \xt^2 + c_2 \yt^2 \ + \ (\xt^2 + \yt^2)^3 \
. $$ This depends only on the two parameters associated to the
quadratic terms, and the analysis of its critical points is simple
enough.

In fact, there is the trivial critical point $$ p_0 = (0,0) \ , $$
always present and stable for $c_1 > 0$ and $c_2 >0$; this is
invariant under the full $G$ group. Then there are some solutions
with both $x$ and $y$ nonzero (the explicit expressions for these
are extremely involved and will not be reported), which are
therefore invariant only under the trivial subgroup made of the
identity alone. Moreover, there are four families of nontrivial
critical points $(\xt,\yt )$, whose existence is limited to ranges
of the parameters $c_1$ and $c_2$. The latter are given by
\begin{eqnarray*}
p_1^\pm &=& \( \pm (- c_1 /3)^{1/4} , 0 \) \ , \\
p_2^\pm &=& \( 0 , \pm (- c_2 / 3)^{1/4} \)  \ .
\end{eqnarray*}

The family $p_1^\pm$ exists for $c_1 < 0$, the family $p_2^\pm$
for $c_2 < 0$. As for their stability, by explicit computations we
obtain that the eigenvalues of the Hessian matrix on $p_1^\pm$ are
given by $\{-8 c_1, -2 c_1 + 2 c_2\}$. Thus in their range of
existence these solutions are stable provided $(c_2 - c_1 ) > 0$;
given that $c_1 < 0$, this is always the case for $c_2 > 0$, while
for $c_2 < 0$ it amounts to the condition $|c_1| > |c_2|$.
Solutions in this family are invariant under $g_y$.

Similarly, by explicit computations the eigenvalues of the Hessian
matrix on $p_2^\pm$ are given by $\{2 (c_1 - c_2), -8 c_2\}$. In
their range of existence these solutions are stable provided $(c_1
- c_2 ) > 0$; given that $c_2 < 0$, this is always the case for
$c_1 > 0$, while for $c_1 < 0$ it amounts to the condition $|c_2|
> |c_1|$. Solutions in this family are invariant under $g_x$.

Thus qualitative information can be obtained by the reduced Landau
potential $\wt{\Phi}$. Should we require to obtain the exact
dependence of the solutions on all the control parameters, we
should invert the change of coordinates $(x,y) \to (\xt,\yt)$;
this inversion should be sought for working by series.

In this case, such an inversion (again non unique) is obtained by
setting
\begin{eqnarray*}
x &=& \xt \ \[ 1 \ - \ (a_1 \xt^2 + a_2 \yt^2) \ + \ \( (3 a_1^2 -
a_3) \xt^4  + (4 a_1 a_2 - a_4
+ 2 a_2 b_1) \xt^2 \yt^2  +
(a_2^2 -
a_5 + 2 a_2 b_2 ) \yt^4 \) \] \ , \\
y &=& \yt \ \[ 1 \ - \ (b_1 \xt^2 + b_2 \yt^2) \ + \ \( (2 a_1 b_1
+ b_1^2 - b_3 ) \xt^4  +(2 a_2
b_1 + 4 b_1 b_2 - b_4) \xt^2 \yt^2 + (3 b_2^2 - b_5) \yt^4 \) \] \
. \end{eqnarray*}

With these, and writing \begin{eqnarray*} \a &=& 3^{3/2} , \ \ \b_1
\ = \ \frac{3 k_1}{2 |c_1|} , \ \ \b_2 \ = \  \frac{ 3 k_3}{2 |c_2|} , \\
\ga_1 &=& \sqrt{3} \( \frac{4 |c_1| (1 - k_4) - k_1^2}{8 |c_1|^2}
\) , \ \ \ga_2 \ = \ \sqrt{3} \( \frac{4 |c_2| (1 - k_7 ) - k_3^2}{8
|c_2|^2} \) , \end{eqnarray*} the solutions in the family
$p_1^\pm$ are given by
$$ (x,y) \ = \ \( \pm \frac{|c_1|^{1/4}}{3^{7/4}} \, \[ \a - \b_1 \sqrt{|c_1|} +
\ga_1 |c_1| \] \ , \ 0 \) \ ;
$$ those in the family $p_2^\pm$ by
$$ (x,y) \ = \ \( 0 \ , \ \pm \frac{|c_2|^{1/4}}{3^{7/4}} \, \[
\a - \b_2 \sqrt{|c_2|} + \ga_2 |c_2| \] \ , \ 0 \) \ . $$

\subsection{Qualitative analysis of Landau polynomials}
\label{sec:redB}

The concrete computational problem in Sect.\ref{sec:redA} was the
determination of the explicit changes of coordinates, i.e. the
expression of $(\wt{x} , \wt{y} )$ in terms of $(x,y)$, and the
inverse transformation expressing $(x,y)$ in terms of $(\wt{x} ,
\wt{y} )$. This can be obtained through the method presented in
the main body of the paper, and would easily produce for the other
examples considered here explicit (and rather lengthy) formulas.

However, as mentioned above, in many cases one would be satisfied
with a qualitative analysis of the Landau potential; that is,
determine which phases are possible and how these change with the
parameters.

Note that it is true that the explicit relation between parameters
in the reduced and in the original Landau potential requires to
explicitly determine the change of variables relating $(x,y)$ and
$(\wt{x} , \wt{y} )$; but it is also true that in a wealth of
physical applications the parameters entering in the (original)
Landau potential are effective ones, determined phenomenologically
by fitting data. So the same approach can be followed directly on
the reduced Landau potential, and we can work directly at this
level.

Needless to say, an analysis of the reduced Landau potential is
much simpler than that of the full (original) one. In this Section
we will shortly indicate how such an analysis can be performed in
the Examples considered above; we will omit Example 1 (and 4),
considered in the previous Section \ref{sec:redA}.

\bigskip\noindent
{\bf Example 2 (continued).} In Example 2, we started from a
potential depending on three parameters associated to the
quadratic part, plus twelve additional ones. All of them can be
eliminated by Poincar\'e changes of coordinates, and recalling the
explicit expression for the $J_a$ we arrive at the reduced
potential \beql{eq:redpotex2} \wt{\Phi} \ = \ c_1 \, x^2 \ + \ c_2
\, y^2 \ + \ c_3 \, x y \ + \ (x^2 + y^2)^3 \ . \eeq The trivial
critical point $(0,0)$ is always present; there are several
branches of nontrivial critical points $(x_*,y_*)$, and we omit
the explicit expressions for these in terms of the $c_i$
parameters. These are invariant under a nontrivial subgroup only
for $c_3 = 0$, in which case the potential \eqref{eq:redpotex2}
reduces to
$$ \wt{\Phi} \ = \ c_1 \, x^2 \ + \ c_2 \, y^2 \ + \ (x^2 + y^2)^3 \ , $$
and we have {\tt (i)} solutions with $y_*=0$ and hence invariant
under the $y$ reflection (for such solutions $x_* = \pm
(-c_1/3)^{1/4}$); and {\tt (ii)} solutions with $x_* = 0$ and
hence invariant under the $x$ reflection (for such solutions $y_*
= \pm (-c_2/3)^{1/4}$). The solutions {\tt (i)} are stable for
$c_1 > 0$ and $c_2 < c_1$, while solutions {\tt (ii)} are stable
for $c_2 > 0$ and $c_1 < c_2$. Thus there is a phase transition
semi-infinite line at $c_1 = c_2$, for both parameters being
positive.

Note that $c_3 = 0$ implies that actually the potential is
invariant not only under the simultaneous reflection in $x$ and
$y$, but separately under reflection in each variable, so that we
are in the frame of Example 1.

\bigskip\noindent
{\bf Example 3 (continued).} In this case, as seen above (but with
a small change of notation), the reduced potential is written as
$$ \Phi \ = \ c_1 J_1 + k_1 J_2 + k_2 J_3 + k_3 J_2^2 + k_4 J_3^2 + k_5 J_2 J_3 + J_1^3 \ . $$
By considering the gradient of this in the orbit space -- that is,
with respect to the $J_i$ variables -- we get immediately that
critical points exist for $c_1 < 0 $ and
$$ J_1 \ = \ \pm \, \sqrt{\frac{- c_1}{3}} \ , \ \
J_2 \ = \ \frac{2 k_1 k_4 - k_2 k_5}{k_5^2 - 4 k_3 k_4} \ , \ \
J_3 \ = \ \frac{2 k_2 k_3 - k_1 k_5}{k_5^2 - 4 k_3 k_4} \ . $$
However, it should be recalled that the three invariants
$J_1,J_2,J_3$ depend on two variables $(x,y)$ (or one complex
variable $z = x + i y$) so that the $\nabla J_i$ are surely not
independent at each point.

Moreover, in this case the symmetry group does not admit any
nontrivial subgroup; it is easily checked that solutions with
other symmetries -- such as reflections in $x$ or $y$ -- only
exist for special values of the parameters and hence do not form a
branch.

\bigskip\noindent
{\bf Examples 5 \& 6 (continued).} The reduced Landau polynomial
is in this cases
\begin{eqnarray*}  \Phi &=& c_1 J_1 + \b_1 J_2 \ + \ \b_2 J_3 \ + \ \b_3 J_2^2 \ + \ \b_4 J_2 J_3 \  + \
\b_5 J_2^3 \ + \ \b_6 J_3^2 \ + \ J_1^6 \ ; \end{eqnarray*}
 By looking at the gradients in terms of the $J_i$ variables, we have
two branches of critical points, i.e.
\begin{eqnarray*}
J_1 &=& \( - c_1 /6 \)^{1/5} \ , \\
J_2 &=& \frac{\b_4^2 \, - \, 4 \, \b_3 \, \b_6 \, \pm \,
\sqrt{\Theta_2} }{12 \, \b_5 \, \b_6} \ , \\
J_3 &=& - \ \frac{\b_4^3 \, + \, 12 \, \b_2 \b_5 \b_6 \, - \,
\b_4 \( 4 \b_3 \b_6 \, \mp \, \sqrt{\Theta_3} \) }{24 \, \b_5 \, \b_6^2} \ ; \\
\Theta_2 &=& 24 \, \b_5 \b_6 \, ( \b_2 \b_4 - 2 \b_1 \b_6 ) \, + \,
(\b_4^2 - 4 \b_3 \b_6)^2 \ , \\
\Theta_3 &=&  \b_4^4 - 8 \b_3 \b_4^2 \b_6 + 24 \b_2 \b_4 \b_5 \b_6 +
16 \b_3^2 \b_6^2 - 48 \b_1 \b_5 \b_6^2  \ .  \end{eqnarray*}

We can have  other solutions at points where the gradients $\nabla
J_i$ are not independent. The matrix built with the gradients of
the three invariants is
$$ M \ = \ 2 \begin{pmatrix} x & y & z \\ x (y^2 + z^2) & y (x^2 + z^2) & z (x^2 + y^2) \\
x y^2 z^2 & x^2 y z^2 & x^2 y^2 z \end{pmatrix} $$
with determinant
$$ \mathtt{Det} (M) \ = \ 8 \ x \, y \, z \ (x^2 - y^2 ) \, (y^2 - z^2) \, (x^2 - z^2 ) \ . $$
Thus the singular sets where gradients of the basic invariants are
not independent is made of the three coordinate axes and of the
six lines bisecting (the positive or negative quadrants of) the
three coordinate planes. One should then consider restrictions of
$\Phi$ to these singular sets; actually due to the inherent
symmetry of the potential, it would suffice to consider just one
case for each type, e.g. just the sets $z=0$ and $z=y$. The
solutions obtained on these singular sets will have a transparent
symmetry and will provide symmetry-breaking solutions.

It should be stressed that these reductions would provide simpler
systems for determination of critical points, i.e. two polynomial
equations in two variables (e.g., choosing the cases mentioned
above, in $x$ and $y$). However, these equations would be of high
degree, degree 10 for the reduction to coordinate axes and degree
eleven for reduction to lines bisecting coordinate planes.

We will not analyze these high degree systems; the symmetry
breakings for this group have been studied in detail in \cite{SGU}
(and also reconsidered in \cite{Gae02}); in particular, Sergienko,
Gufan and Urazhdin considered in detail the different type of
phase transitions occurring in this case, and the reader is
referred to their work for a detailed (quantitative and not just
qualitative) analysis.

\section{Discussion}
\label{sec:disc}

We have so far shown that all terms in the range of the
homological operator $\L_0$ can be eliminated by a suitable
sequence of Poincar\'e transformations, and shown how one can
proceed in practice to obtain this.

We will now briefly discuss the advantages, together with the
limitations and some possible extensions of our approach.

\subsection{Advantages of the method}

In studying the behavior of (the extremal points of) the Landau
polynomial $\Phi$ when the parameters appearing in it are varied,
one is usually faced with a formidable task, just due to the high
number of these parameters. In fact, the general approach should
go through a study (often possible only via a numerical approach,
in particular for high $N$) of the critical point of $\Phi$,
exploring a high dimensional parameter space.

The advantage of the method proposed here, which is just a
reformulation of the Poincar\'e approach to the study of dynamical
systems around an equilibrium point (or other known solutions),
lies in that the number of parameters, and thus the dimension of
the space to be explored, is reduced. This reduction can in fact
be quite substantial, as we have seen in some of the Examples
considered through Section \ref{sec:Examples}. In fact, in Example
1 (and Example 4) we passed from nine to two parameters, in
Example 2 from fifteen to three parameters, in Example 3 from ten
to six parameters, in Example 5 and 6 from from twenty-two to six
parameters.

Thus the effectiveness of the method depends on the group
(representation) one is considering. Moreover, while a problem
depending on two parameters can be analyzed, a problem depending
say on six parameters is still extremely hard to analyze; so
obviously the present method provides in general a step forward,
but not a full solution.

We will now pass to consider several other limitations of the method

\subsection{Varying parameters}

First of all it should be stressed that we have worked with a
given Landau polynomial, i.e. with {\it fixed} values of the
parameters entering in it (the coefficients of the Landau
polynomial). These parameter -- or at least some of them -- will
in general depend on the external ``control'' parameter, i.e. the
physical ones: temperature, pressure, magnetic field, etc; and
indeed the Landau parameters {\it have } to change with the
physical ones for a phase transition to take place. Thus some
extra care is needed if we want to work on a full interval of
values of the control parameter(s).

In particular, one is often interested in (the vicinity of) phase
transitions; in this case the coefficients of the polynomial $\Phi
(x)$ not only depend on external control parameters $\la$, but at
phase transition necessarily pass through critical values.

The discussion given so far should be modified if we want to
consider not just given fixed values of the parameter(s) but a
full range of values, including in particular critical ones.

Let us consider, for ease of discussion, a single control
parameters $\la \in \La \sse \R$; and let $\la_0 \in \La$ be a
critical value. If we want to describe a small but finite interval
$\La_0 \ss \La$, we have to require that the near-identity changes
of variables considered in previous sections are defined uniformly
in $\La_0$. In particular, if we want to consider an interval
which includes the critical point, e.g. $\La_0 = [\la_0 - \eps ,
\la_0 + \eps]$, this would mean requiring that these changes of
variables are well defined also at $\la = 0$. Note that the
changes of variables considered in Section \ref{sec:ExII} do not
in general pass this criterion: e.g. many of them are not allowed
when $c_1 = 0$.

It should be stressed that, as mentioned in Remark 1 above, this
is just inherent to the method. In fact, the vanishing of $c_1$
means the vanishing of the quadratic part of the Landau potential,
on which all of the Poincar\'e procedure is based. Note also that
this makes perfect physical sense: we cannot expect to have
results uniform in $\la$ over an interval which includes a phase
transition.

The conclusion is that special care must be taken (as also rather
obvious physically) if we want to consider reduction of the Landau
polynomial over a full range of parameters, and in particular the
allowed reduction is (in general, severely) limited if this range
includes critical values (actually, one should avoid these). On
the other hand, our method can give a simplified description of
the outcome of a phase transition, analyzing the simplified
potential for values of the parameters higher or lower than the
critical ones.

\subsection{Small denominators}

The generating functions for the Poincar\'e near-identity changes
of variables are obtained as solutions of the homological
equation. As seen quite clearly in the Examples, and  as is
specially clear once the $Q$ matrix (or its semisimple part) has
been set in diagonal form, this involves inversion of a matrix and
thus introduces some denominators.

When the latter vanish, the transformation is not defined and
hence the reduction turns out to be impossible. But even when the
denominators are nonzero, some care should be taken if they are
small.

In fact, our approach is based on a series expansion; for this to
make sense it is needed that the terms of different orders have a
size which correspond to their order (that is, that the series is
well ordered). If the expansion parameter (roughly speaking, the
distance from the critical point) is $\eps$ and we perform a
change of variables in which the involved denominator is larger
than $\eps^{-1}$, then terms which are apparently of order
$\eps^k$ will actually be of lower orders, and the series is not
well ordered any more. In other words, the series expansion gets
not justified in this case.

Thus one should check the appearance of these small denominators;
they will in general make that the resulting change of coordinates
are well defined only within a certain radius of convergence, and
the computation will have physical relevance only if the minima of
the Landau polynomial (i.e. the physical state) lies within this
convergence region.

It should be mentioned that some way to partially escape this
problem is well known in dynamical systems. In fact, the small
denominators will appear only when attempting to eliminate terms
which correspond to near-resonances, i.e. such that $\la_\a \cdot
k_\a \simeq 0$, see eq.\eqref{eq:res}. It is thus possible to
circumvent them by simply renouncing to eliminate near-resonant
terms.

In more formal terms, this is obtained by ``detuning the
resonance'' \cite{Ver}: we write $\la_\a = \s_\a + \eps^2
\eta_\a$, and consider $\s_\a$ as the eigenvalues of $P$, while
the difference $\la_\a - \s_\a$ is considered as a perturbation
term \footnote{It is necessary to consider this as being second
order in $\eps$ for the method to be viable \cite{Ver,Puc}.}, to
be included between higher order terms in the Landau expansion.

We will not discuss this approach here (see \cite{Puc} for a
recent overview), but it is worth mentioning that it gave
extremely satisfactory results in explicitly computing quantum
levels of molecules up to near the dissociation threshold
\cite{Joy1,Joy2}.

\subsection{Non orthogonal action}

We have assumed the group $G$ acts in $M = \R^n$ by an orthogonal
action. Unfortunately this is not always the case in concrete
applications; albeit in principles (by Palais-Mostow theorem, see
e.g. \cite{AbS,MZZ}) one can always reduce to an orthogonal
action, this goes through dimension increase and/or modification
of the metric. This means that in practice the method can become
much more involved and less computationally convenient.

In particular, it is known that Landau theory for liquid crystals
\cite{deG,Vir} requires a description in terms of a tensorial
order parameter of second order; the natural group action on this
is not fitting simply in the framework considered here, and will
be discussed elsewhere.

\subsection{Further reduction}

In our discussion we have considered the result of changes of
variable on terms of the same order as the generating function,
without describing in detail the higher order effects.

Actually, the reduction procedure can be iterated -- restricting
to generating functions in the kernel of $\L_0$, so not to change
terms which have already been reduced -- using higher order
effects; the latter are basically controlled by using the
Baker-Campbell-Haussdorff formula. This is known (under different
approaches) in the framework of dynamical systems as ``further
normalization'', and we will just refer the interested reader to
e.g. \cite{CGs,Gae02,RNF} and references therein.

\subsection{Dynamics and Landau-Ginzburg}

In standard Landau theory, the equilibrium state $x$ of the physical system is
described by the minima of the Landau potential $\Phi (x)$; this
description is inherently static. One can also provide, in nearly
the same terms, a dynamical description; the time evolution of the
state $x(t)$ of the system is then described by $\dot{x} = -
\nabla \Phi (x)$. Needless to say this agrees with standard Landau
theory if we look at asymptotic solutions.

In this case the equations of motion also include a $\dot{x}$
term, and our computations for the effect of a change of variables
on the equations have to be changed accordingly. Once again we
will defer a detailed account of this modification of our
approach, and just refer the reader to the equivalent treatment
given in the dynamical systems framework (by construction, one
would be interested only in the time evolution of invariants); see
e.g. \cite{DLan}.

A well known extension of Landau theory is provided by
Ginzburg-Landau theory; here the order parameter is a local
function on spacetime, and the theory is described by a (gauge)
invariant functional. Michel theory can be extended to this
framework \cite{GaMor}, and it has thus to be expected that our
approach works also here. \footnote{There will be a severe
reduction of the regularity of the functional upon reduction to
the orbit space \cite{Ball}; luckily this reduction does not take
place if one works with $C^\infty$, or analytic, functionals. I am
grateful to prof. Ball for pointing out these facts.}

\section{Conclusions}
\label{sec:concl}

In the Landau theory of phase transitions, one considers an
effective potential $\Phi$ whose symmetry group $G$ and degree $d$
depend on the Physics of the system under consideration.

One should consider as $\Phi$ the most general $G$-invariant
polynomial of a certain degree $d$. When such a $\Phi$ turns out
to be too complicate for a direct analysis, it is essential to be
able to consider a simplified potential $\^\Phi$ giving raise to
the same behavior as the original one.

Here we have described in detail a reduction procedure based on
classical Lie-Poincar\'e theory; this just considers changes of
variables, defined locally. Thus, it expresses the {\it same}
potential $\Phi$ in {\it different} coordinates.

In many cases one is satisfied with analyzing the behavior of the
Landau potential for fixed values (near the transition point) of
the control parameter(s); in these cases our method is specially
effective. In other cases one wants to be able to ``follow'' the
critical points of the Landau potential as the control
parameter(s) is (are) changed over a range $\La$; in this case one
has to require the change of variables required by our method are
uniformly defined in all of $\La$, which poses serious limitations
on the applicability of the method.

We stress that our discussion does not just provide a proof of the
fact one can consider a reduced potential of the form described in
detail in previous sections; it also gives a constructive
algorithm to make completely explicit computations.

We have shown this by a number of explicit examples; these
included in particular groups describing the symmetry of
isotropic, non-isotropic and chevron-shaped nematics.


Finally, we would like to stress that in this paper we consider
just changes of coordinates: we eliminate terms by choosing
suitable coordinates, but we are {\it not } changing the physical
potential. On the other hand, in Landau theory one allows changes
of the potential, provided these do not alter its qualitative
behavior.

Thus, we are {\it not } considering the most general
transformation of $\Phi$ allowed by Landau theory. On the other
hand, our reduction amounts to a change of variables, requires
only to solve linear equations, and is completely algorithmic; it
can thus be easily implemented, maybe resorting to a symbolic
manipulation language in order to perform the algebraically
complex (albeit conceptually simple) required computations.
Further analysis -- maybe with an actual change of the physical
potential, and based on physical considerations rather than on
mathematical manipulations -- can then be applied on the
simplified form of the potential thus obtained.

\section*{Acknowledgments}
This work was triggered by participation in the Newton Institute
workshop on ``Symmetry, bifurcation and order parameters''; I
thank D. Chillingworth for his invitation there, and several
participants for interesting remarks.
The first version of the paper was written while
visiting LPTMC (Paris-Jussieu); I thank M. Barbi for her
hospitality. My research is partially supported by MIUR-PRIN
program under project 2010-JJ4KPA.

\end{document}